\documentclass[5p,times]{elsarticle}

\usepackage{lineno,hyperref}
\usepackage{subfig}
\usepackage{upgreek}
\usepackage{amsmath, amsthm}
\usepackage{gensymb, textcomp}
\usepackage{graphicx}
\usepackage{multicol}
\usepackage{placeins}

\modulolinenumbers[5]

\journal{Journal of \LaTeX\ Templates}

\biboptions{numbers,sort&compress}
\begin{document}

\begin{frontmatter}

\title{A study in using MICROMEGAS to improve particle identification with the TAMU-MDM focal plane detector}

\author[a,c]{Spiridon A.}

\author[b]{Pollacco E.}
\author[a]{Saastamoinen A.}
\author[a]{Dag M.}
\author[a]{Roeder B.}
\author[a]{Tribble R.}
\author[c]{Trache L.\corref{mycorrespondingauthor}}
\cortext[mycorrespondingauthor]{Corresponding author: livius.trache@nipne.ro}
\author[c]{Pascovici G.}
\author[d]{Mehl B.}
\author[d]{de Oliveira R.}

\address[a]{Cyclotron Institute, Texas A \& M University, College Station, TX, 77843-3366}
\address[b]{IRFU, CEA, Universit\'{e} Paris-Saclay, F-91191, Gif-sur-Yvette, France}
\address[c]{National Institute of Physics and Nuclear Engineering, Bucharest-Magurele, RO-077125, Romania}
\address[d]{CERN (EP-DT-ED), Geneva, Switzerland}

\begin{abstract}
A MICROMEGAS detection amplifier has been incorporated into the design of the TAMU-MDM focal plane detector with the purpose of improving the energy resolution and thus, the particle identification. Beam tests showed a factor of 2 improvement over the original design, from 10-12\% to 4-6\%, for ions with A$\le$32 at E/A $\sim$ 10--20 MeV.
\end{abstract}


\end{frontmatter}

%\linenumbers

\section{Introduction}
\emergencystretch\linewidth
The Multipole-Dipole-Multipole (MDM) spectrometer at the Cyclotron Institute, Texas A \& M University has been in use for over 25 years, since it was brought from the University of Oxford in 1992 \cite{MDM1986} together with the focal plane detector \cite{Oxford1986,Youngblood1995}. Since then, numerous experiments have been performed with this beamline for giant resonance studies, as well as for astrophysical reaction rate studies, among others.

The MDM focal plane detector, also called the ``Oxford detector'', has been used in particular to study elastic scattering and transfer reactions for the determination of astrophysical reaction rates using the Asymptotic Normalization Coefficient (ANC) method \cite{LT2000,Mukh1997,LT2003}. The detector provided position information for raytrace reconstruction and energy loss signals for particle identification. For these experiments, it was important to be able to separate A and A+1 nuclei and the Oxford detector has done this successfully for particles with masses up to and including A=22 \cite{Tariq2010}. A study of the reaction $^{13}$C($^{26}$Mg,$^{27}$Mg)$^{12}$C showed that this was at the limit of the detector, or beyond it, in terms of its particle identification (PID) capabilities.

This limitation sparked the idea of modifying the Oxford detector to increase its resolution in measuring energy loss. A contributing factor to this was also the ongoing facility upgrade at the Cyclotron Institute intended to provide unstable re-accelerated beams. 

\begin{figure}[ht!]
	\centering
	\includegraphics[width=0.9\linewidth]{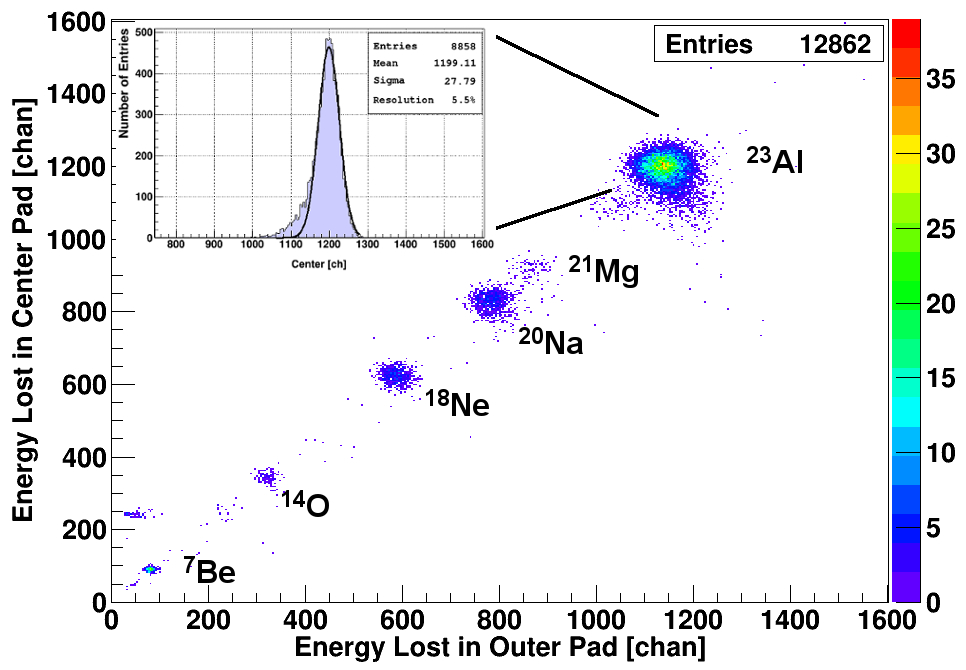}
	\caption{2-D histogram measured using AstroBox showing the energy losses of a $^{23}$Al beam and its contaminants. The plot gives energy loss in the central pad (Y axis) versus the energy loss detected by one of the outer pads (X axis). The insert represents a Y-axis projection of the $^{23}$Al data giving the energy resolution.}
	\label{fig:AB23Al}
\end{figure}

The idea of how to improve the energy resolution of the Oxford detector came from a previous study that involved building a detector for low-energy protons from beta-delayed proton decay. This instrument, called AstroBox \cite{AB1_2013}, used Micromegas technology \cite{MuO1996} to not only successfully measure proton energies as low as $\sim$100 keV without being overwhelmed by the beta background, but as shown in Fig. \ref{fig:AB23Al}, it was also able to detect heavier ions with very good separation for a good range of mass numbers.

Given the positive results obtained with AstroBox and the relatively easy operation of the Micromegas, it was decided that modifying the Oxford detector to include Micromegas for energy detection would be faster, less costly and with the potential to be more successful than any other option for an upgrade. Preliminary reports on this upgrade project can be read in \cite{Spiridon2016} and \cite{SpiridonPhD}.
%
%-----------------------------------
%
\section{The original detector}
\begin{figure*}[ht!]
	\begin{center}
		\subfloat[]{\includegraphics[width=0.53\linewidth]{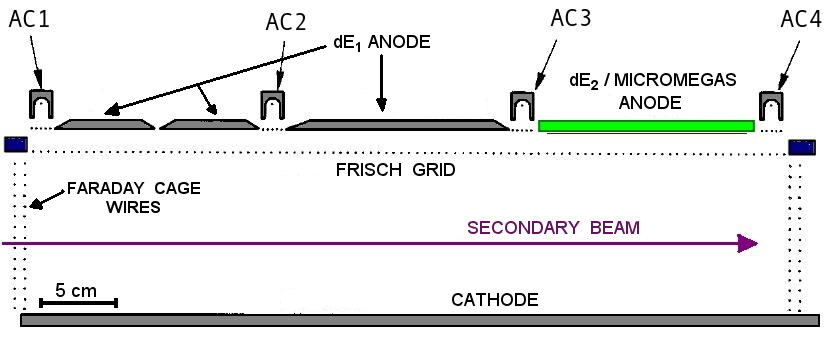}}
		\hspace{0.5 cm}
		\subfloat[]{\includegraphics[width=0.3\linewidth]{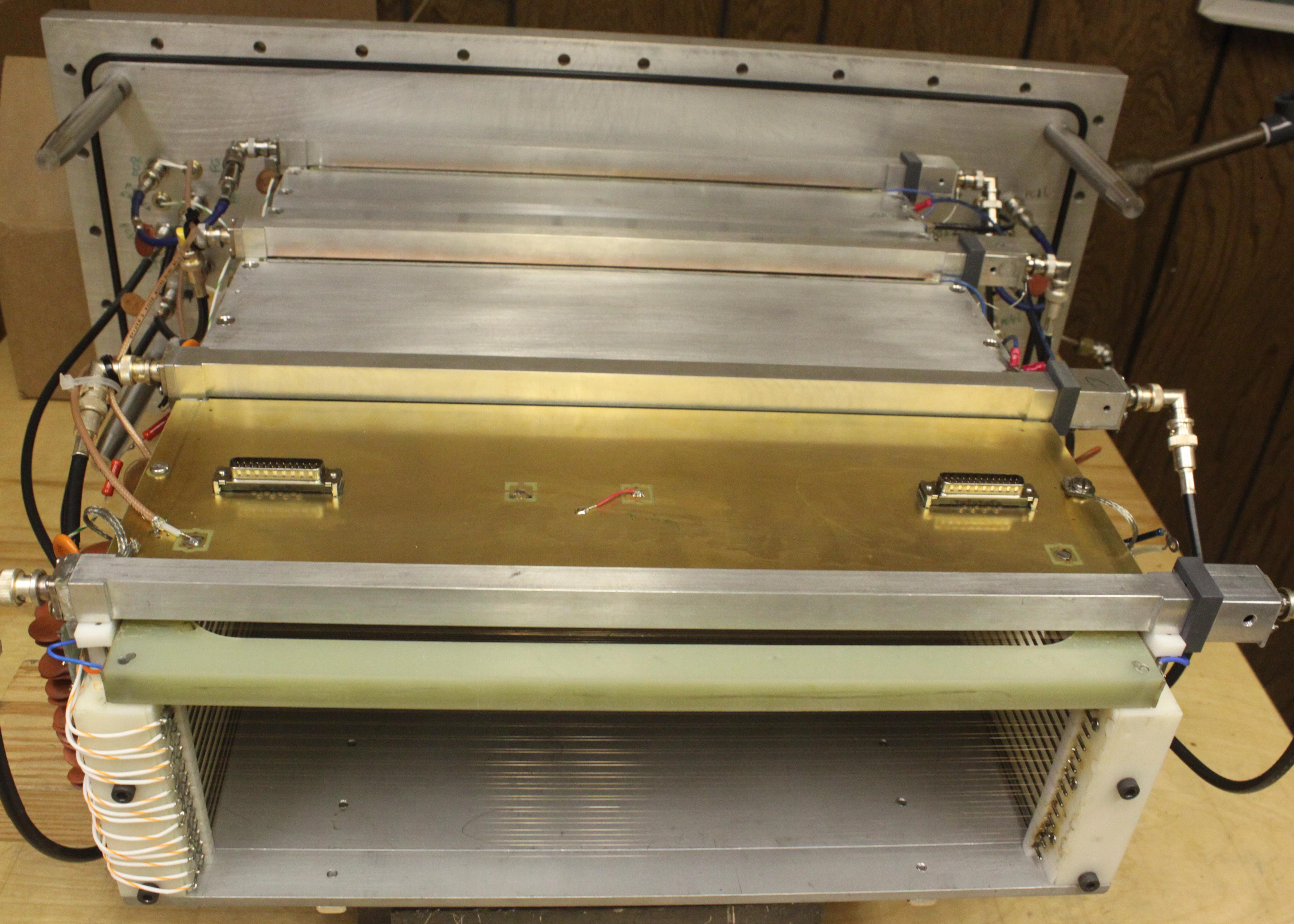}}
	\end{center}
	\caption{(a) Schematic drawing of the inside components of the Oxford detector showing the position of the new Micromegas anode. (b) Photograph taken by authors showing the inside components of the Oxford detector and the mounted Micromegas anode.}
	\label{fig:UpgOxf}
\end{figure*}
The Oxford focal plane detector is a gas-filled gridded ionization chamber with 4 resistive avalanche counters (ACs) and 3 aluminum anodes. These ACs are used to measure position at four depths inside the detector to determine the angle of the particle track for RAYTRACE \cite{CodeRaytrace} reconstruction. The anodes are used to determine the energy lost in the gas and are connected in a manner that gives 2 energy loss signals. Isobutane gas is used at pressures between 30 and 200 Torr, depending on the nuclei studied. A Frisch Grid (FG), along with fourteen electrodes (thin bars) going around the four sides, form a Faraday cage that ensures field uniformity inside the detection region \cite{Oxford1986}. Two photomultipliers (PM) are coupled to a plastic scintillator plate and attached to the back of the detection chamber. The scintillator is used to stop the nuclei and measure their residual energy. The PM signals are also used to trigger the data acquisition system \cite{Youngblood1995}.

In summary, the Oxford detector provided 11 output signals: 8 for position determination, 2 for energy loss (called \textbf{dE1} and \textbf{dE2}) and one for the residual energy (label \textbf{PM}). The specific gas pressure and scintillator thickness, as well as the voltages on the various elements of the Oxford detector are chosen specifically for each experiment, with the goal of having the secondary beam particles of interest pass through the gas and stop in the scintillator and be detected with optimal resolution. In these circumstances, energy resolutions for \textbf{dE1} and \textbf{dE2} varied between 10\% and 17\% depending on gas pressure (the lower the pressure, the poorer the resolution). Additionally, \textbf{dE2} was consistently worse than \textbf{dE1} because the signal is smaller (shorter path of travel) and the straggling effect from the particle passing through the previous sections becomes more significant. Moreover, for gas pressures below 30 Torr, the \textbf{dE2} signal tends to have a significantly lower signal to noise ratio (S/N) making it unusable.
%-----------------------------------------------
%
\section{The MICROMEGAS upgrade}
The upgrade of the Oxford detector was focused on improving the energy loss detection with Micromegas by obtaining relatively high gains and reaching a higher signal to noise ratio. This work was undertaken when Micromegas technology was practically not used in Nuclear Physics. The modified section of the focal plane detector consists of two regions. Particles pass through a drift gap (several cm across), causing ionization in the gas. The positive ions are collected by the cathode, while the electrons drift though the Frisch grid and enter the Micromegas. The electrons are focused through the stainless-steel mesh of the Micromegas with high efficiency and are subsequently amplified in the gap via an avalanche mechanism. With appropriate electric fields in the two regions, this technology has been shown to provide gains as high as 10$^{5}$ \cite{MuO1996}. In essence, the Micromegas component acts as an amplifier for the ionization signal created in the drift region.

The main concern about using this technology was that such a detection scheme, combining Micromegas with a gridded ionization chamber, had not been used before. A lesser concern was that our previous knowledge (see Ref. \cite{AB1_2013} on AstroBox) of operating the Micromegas lay close to the atmospheric pressure regime and not the low pressures ($\leq$ 200 Torr) needed for heavy ions in the Oxford detector. 

Considering these initial unknowns, the upgrade had to be reversible. If the modifications were not successful, it was important for us to be able to revert to the original design without losing significant experimental time. The simplest method to achieve this was to replace the \textbf{dE2} anode (Fig. \ref{fig:UpgOxf}) with a Micromegas anode of identical geometry.

\begin{figure}[ht!]
	\centering
	\includegraphics[width=0.95\linewidth]{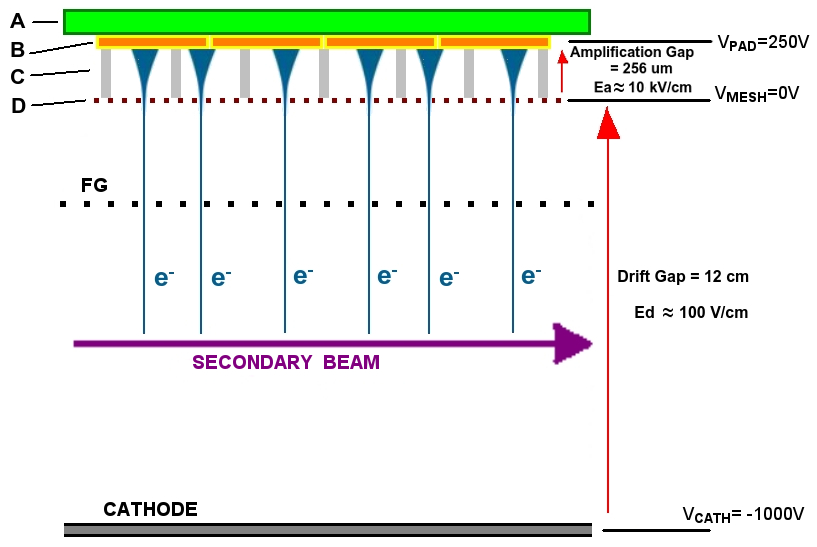}
	\caption{Schematic of the Micromegas detector. Sizes are not to scale. Micromegas elements are labeled with capital letters: A-PCB, B-anode pads, C-insulating pillars and D-micromesh. The electron sheet is also indicated, in blue color and with `e$^{-}$'.}
	\label{fig:MuO-scheme}
\end{figure}

The new anode consists of a circuit board (labeled A in Fig. \ref{fig:MuO-scheme}) printed with gold-plated copper anode pads (labeled B in Fig. \ref{fig:MuO-scheme}).The PCB is 6 mm thick to give close to perfect planarity. Each pad is 32.5 mm deep (along the beam) and 44 mm wide (across the beam) giving a total of 28 pads (4 rows of 7 pads) and forming a detection area of 13.5 cm by 30.9 cm. Below the pads is a micromesh (labeled D in Fig. \ref{fig:MuO-scheme}) made of stainless steel inter-woven wires with diameter of 18 $\upmu$m and a pitch of 63 $\upmu$m. 

The electrons’ transparency (95\%) through the mesh \cite{MuO1996} is attained by reaching an optimized field ratio between the drift and avalanche zones. Bias on the anode leads to a field $\sim$10 kV/cm and yields an avalanche amplification region in the gap of 256 $\upmu$m. The mesh is supported at a uniform distance by resin pillars (labeled C in Fig. \ref{fig:MuO-scheme}), with diameter of 0.3 mm and pitch of 5 mm. The 256-micron gap allows a relatively high gain at low pressures by giving the electrons a longer path to develop the avalanche. When the Micromegas is mounted on the Oxford detector plate, the mesh creates a drift gap with the cathode of 12 cm. Field uniformity in this region is ensured by the Oxford detector Faraday cage. Typical bias voltages are shown in Fig. \ref{fig:MuO-scheme}. For the anode pads, the bias was varied for optimization.

Initially, we wanted a single large area of detection, like the previous \textbf{dE2} anode. However, in that case its capacitance would have been $\sim$2 nF, which would have reduced the signal to noise ratio. The current pad dimension is the largest that could be used while keeping a reasonable S/N ratio ($\sim$300:1). Another concern was that the charge created by particles with high Z over the entire surface would be large, even at low voltages, and would trigger sparking and detector breakdown. These
effects, although present, were rendered insignificant by appropriate tuning.

The 28 individual signals are routed through the internal circuit of the PCB to two DSub-25 connectors and from there to the vacuum-air feedthroughs. Two Mesytec MPR16 preamplifiers are directly connected to the feedthrough flanges in order to minimize noise. The shaping of the signals was done with 2 Mesytec MSCF16 modules and the data acquisition trigger was given by the PM signals.
\section{Tests and results}
The Oxford detector upgrade was tested with a variety of beams. Specifically, there were 6 beams used: $^{16}$O, $^{22}$Ne, $^{26}$Mg, $^{27}$Al, $^{28}$Si and $^{32}$S. In each case, the beam energy was approximately 12 MeV/A and the main target was $^{197}$Au. The gas choice of isobutane was not changed throughout the tests. The Micromegas element was the same throughout all the experiments, with a 256 $\upmu$m gap.

To characterize the performance of the Micromegas, the elastically scattered beam was collimated with
a narrow slit (0.1\degree{} wide). The Micromegas response was plotted in individual pad histograms containing the raw data. Throughout this paper, individual pads will be referred to according to their row and column, for ex. R1-C1 represents the pad in row 1 and column 1.
%
%---------------------------------
%
\subsection{Efficiency}
\begin{figure}[!ht]
	\centering
	\includegraphics[width=\linewidth]{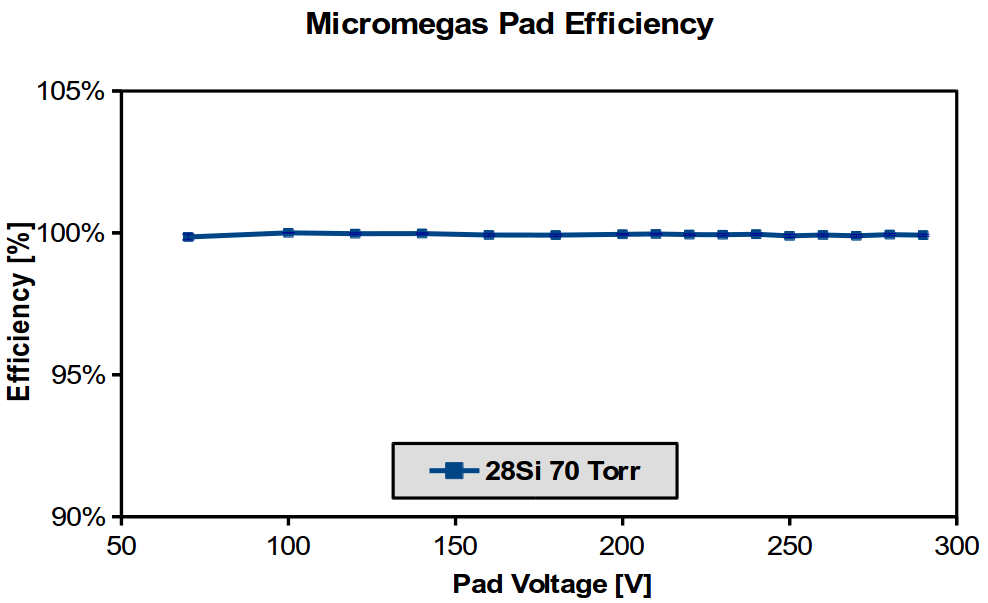}
	\caption{Detection efficiency of pad R1-C4 as a function of pad bias voltage for $^{28}$Si particles in isobutane at 70 Torr. The Y-axis error bars indicate statistical uncertainties.}
	\label{fig:4.EffTest}
\end{figure}

The detection efficiency was evaluated as the ratio between the counts recorded by the Micromegas pads and the counts detected by \textbf{dE1} (the ionization detection region of the Oxford detector). Noise related counts are excluded. This ratio can be seen as a relative efficiency since it depends on the performance of the \textbf{dE1} component of the Oxford detector. Fig. \ref{fig:4.EffTest} shows the efficiency of pad R1-C4 as a function of the pad bias voltage for elastically scattered $^{28}$Si particles passing through isobutane at 70 Torr. It can be seen that the efficiency is close to 100\% for the entire range of bias voltages. This evaluation was done for all 28 pads with similar results. The detection efficiency across the Mcromegas anode was found to be between 99.5\% and 100\%.
%
%-----------------------------------------
%
\subsection{Linearity}
\begin{figure*}[!ht]
	\begin{center}
		\subfloat[]{%
			\includegraphics[width=0.25\textwidth]{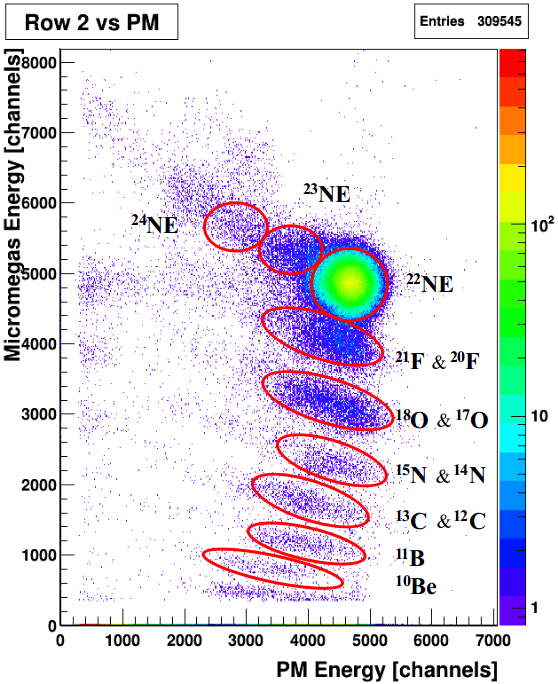}
		}%
		\hspace{0.7 cm}
		\subfloat[]{%
			\includegraphics[width=0.6\textwidth]{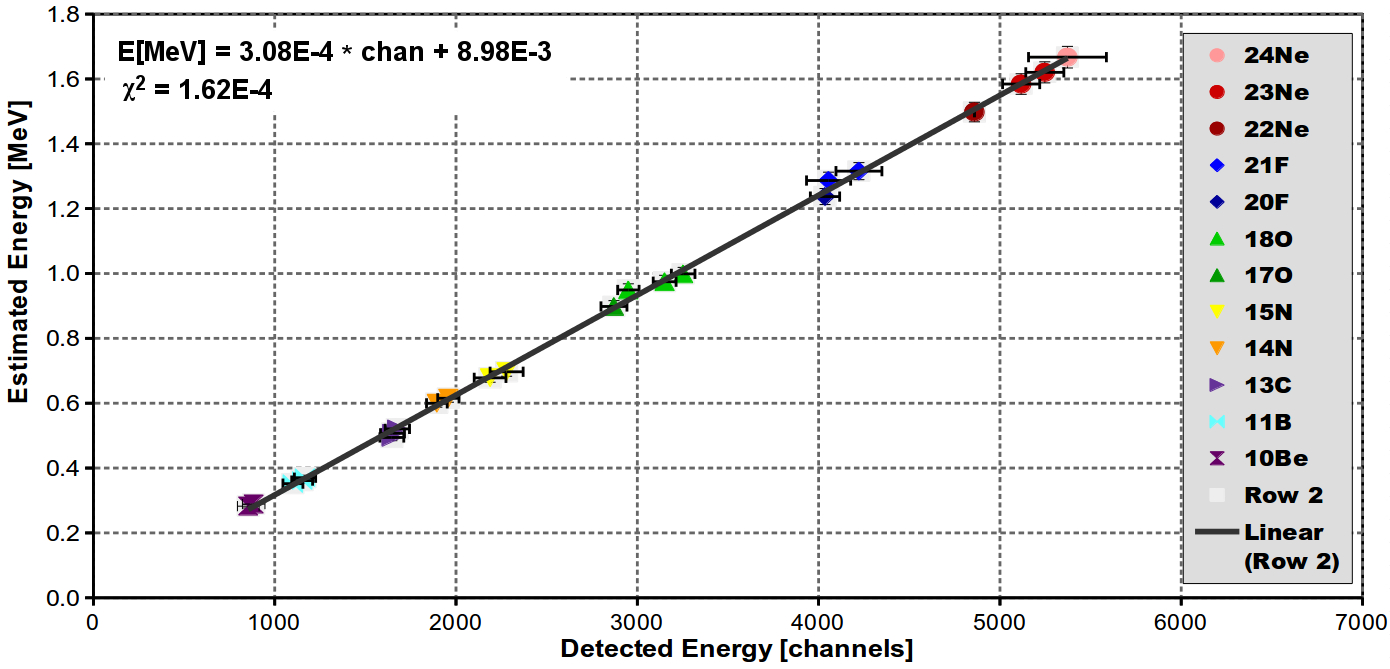}
		}
	\end{center}
	\caption{(a) Micromegas row 2 energy versus residual energy for a $^{22}$Ne beam at 12 MeV/A at a pressure of 30 Torr. (b) Linearity plot for the total energy loss in Row 2. The X-axis error bars indicate statistical uncertainties. The Y-axis error bars indicate the uncertainty in the energy loss estimation.}
	\label{fig:Linear}
\end{figure*}

In order to observe the linearity of the Micromegas response, it was necessary to have different amounts of energy deposited in the gas. The method to study this characteristic involved using a $^{22}$Ne beam at 12 MeV/A on a $^{13}$C target (100 $\upmu$g/cm$^{2}$). The result was a cocktail of reaction products, as can be seen in Fig. \ref{fig:Linear}, (a), which shows a 2-D histogram with row 2 response on the Y-axis and stopping energy on the X-axis.

The gas pressure in this specific case was 30 Torr. The various reaction products are indicated in the figure. The circled events were separated with a software gate and fitted with Gaussian distributions. Those data were then plotted versus position in the focal plane. Ground states and specific excited states were then determined leading to an estimate of energy loss in MeV using TRIM \cite{SRIM}. In each case, the response of the Micromegas was also determined in channels by fitting the corresponding peaks. Fig. \ref{fig:Linear}, (b) shows the estimated energy loss on the Y-axis and the response of Row 2 of the new anode on the X-axis. It can be seen that the Micromegas linearity is quite good across the investigated range (the normalized $\chi^2$ of the fit was 1.62$\cdot$10$^{-4}$).
%
%--------------------------------------------------------------------
%
\subsection{Gain}
The gain of the Micromegas was determined relatively with the Oxford setup and as such, only its variation with various parameters was tested. We defined it as
\begin{equation}
Gain = \frac{N_{\text{total electrons}}}{N_{\text{ionization electrons}}} \, ,
\end{equation}

\hspace{\parindent}where $N_{\text{ionization electrons}}$ represents the average number of electrons produced in the initial ionization process and was determined from the ratio between the energy lost in the gas and the average energy needed to produce an ion pair, $\frac{E[eV]}{w}$. This number represents a rough estimate as not all the energy loss produces ion-pairs. The average energy, $w$, for isobutane is $\sim$23 eV/electron-ion pair \cite{Leroy2015} and takes into account the fact that some pairs recombine. 

The total number of electrons, $N_{\text{total electrons}}$ collected by the Micromegas anode was defined as the ratio, $\frac{Q[\text{pad}]}{e}$, of the charge collected on each pad to the electron charge. To determine the charge Q we have used a calibration procedure that is not detailed herein (see Ref. \cite{SpiridonPhD}). The dependence on the amplification field was checked by changing the Micromegas anode bias, from 0 to $V_{max}$. The maximum voltage, $V_{max}$, that could be applied depended on the energy loss of the ionizing particle. Given a range of pad voltages of $V_{pad}$=100--300 V, amplification fields of up to 12 kV/cm were obtained without breakdown. In all cases, the ADC range limit was reached before the gas breakdown limit. Similarly, the gain variation with pressure and Z number of the ionizing particles were tested. 

\begin{figure}[!ht]
	\centering
	\includegraphics[width=0.9\linewidth]{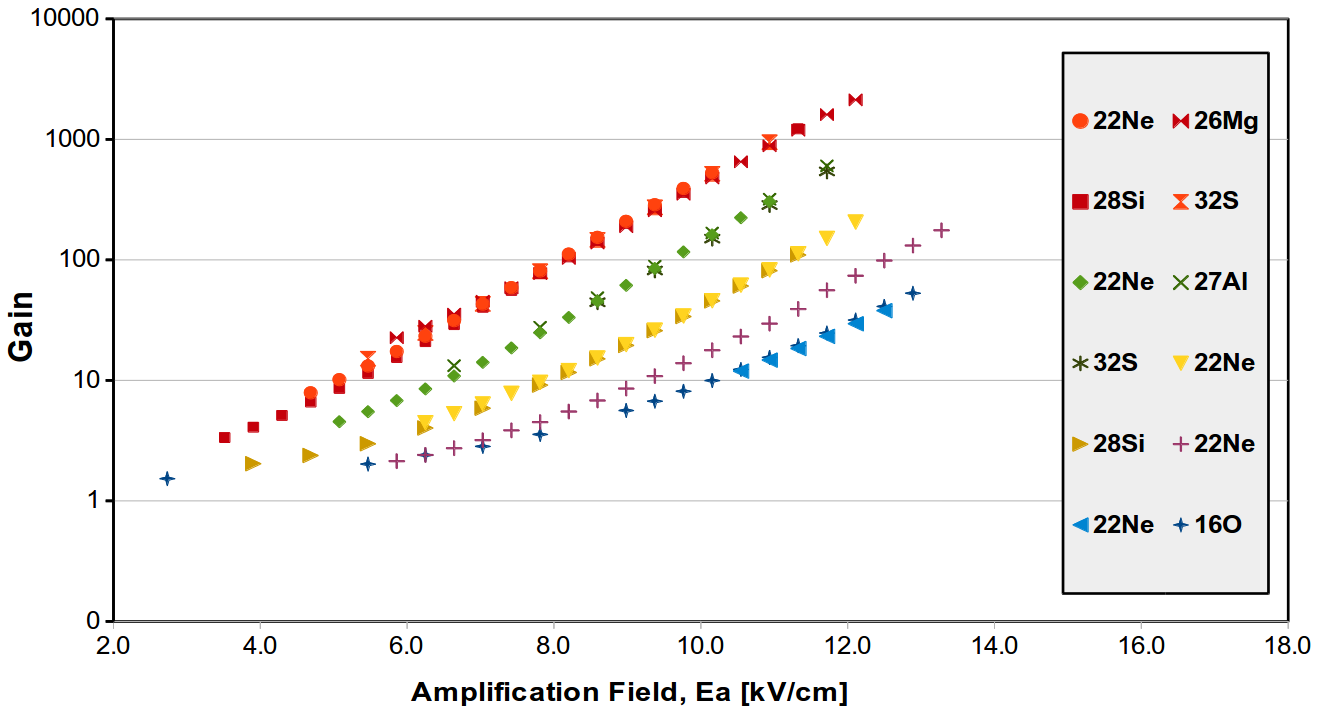}
	\caption{Micromegas gain curves for all the ionizing particles used in the testing. The different gas pressures are color coded (Torr): red=30, green=50, yellow=70, purple=85 and blue=100. The Y-axis error bars indicate statistical uncertainties but are too small to be visible.}
	\label{fig:GainAll}
\end{figure}

Specifically, each of the 6 beams mentioned in the beginning of the section was collimated with the narrow slit and elastically scattered off the $^{197}$Au target. The scattered beam was detected with the Micromegas and the resulting data are shown in Fig. \ref{fig:GainAll}. 

The different colors of the curves indicate the pressure values, as noted in the figure caption. The trend indicates an increase in gain with decreasing pressure for the same amplification field. In addition, we found that data points taken at the same pressure fall approximately on the same curve, independent of the type of ionizing particle, which also agrees with expectations. While factors greater than 10$^{3}$ are desirable in other cases, the gain results obtained in this work are high enough to ensure good signal to noise ratio for this application.
%
%
%----------------------------------------------
%
\subsection{Energy resolution}
\begin{figure}[!ht]
	\centering
	\includegraphics[width=\linewidth]{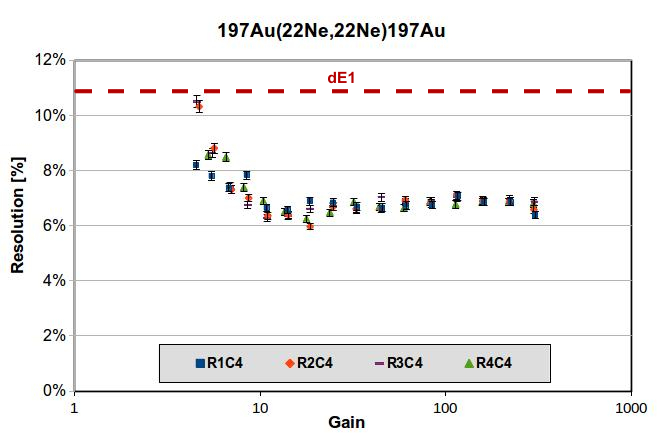}
	\caption{Resolution variation with gain for the central pad in each row of the Micromegas anode, with a beam of $^{22}$Ne at 50 Torr pressure. The Y-axis error bars indicate statistical uncertainties. The red dashed line indicates the \textbf{dE1} resolution for this case, for comparison purposes.}
	\label{fig:ResNe50}
\end{figure}

Since the focus of this upgrade is the energy resolution, this was studied for different gain/bias voltages and gas pressures. We defined the relative resolution for each pad as the FWHM of the energy loss peak. As an example, Fig. \ref{fig:ResNe50} shows the energy resolution variation with gain for the $^{22}$Ne beam, for 50 Torr pressure. The shape exhibits the threshold region between proportionality and amplification. From this figure, for this particular beam and pressure, the best setting to run at was with pad bias $V_{pad}$=260 V (gain $\approx$ 150), both in terms of resolution as well as signal strength.

\begin{figure}[!ht]
	\centering
	\includegraphics[width=\linewidth]{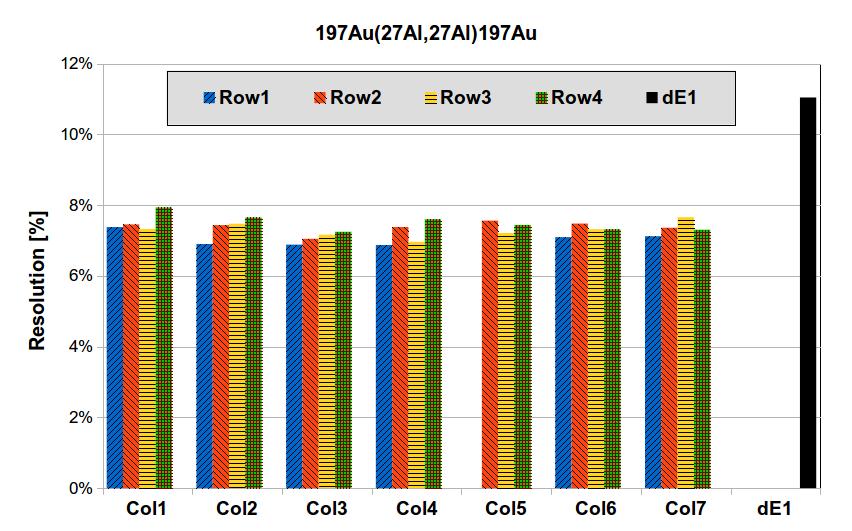}
	\caption{Individual pad resolutions for \textsuperscript{27}Al at 50 Torr pressure. The solid black bar represents the \textbf{dE1} resolution for this beam and pressure and was added for comparison purposes.}
	\label{fig:ResAll-Al50}
\end{figure}

\begin{figure}[!ht]
	\centering
	\includegraphics[width=\linewidth]{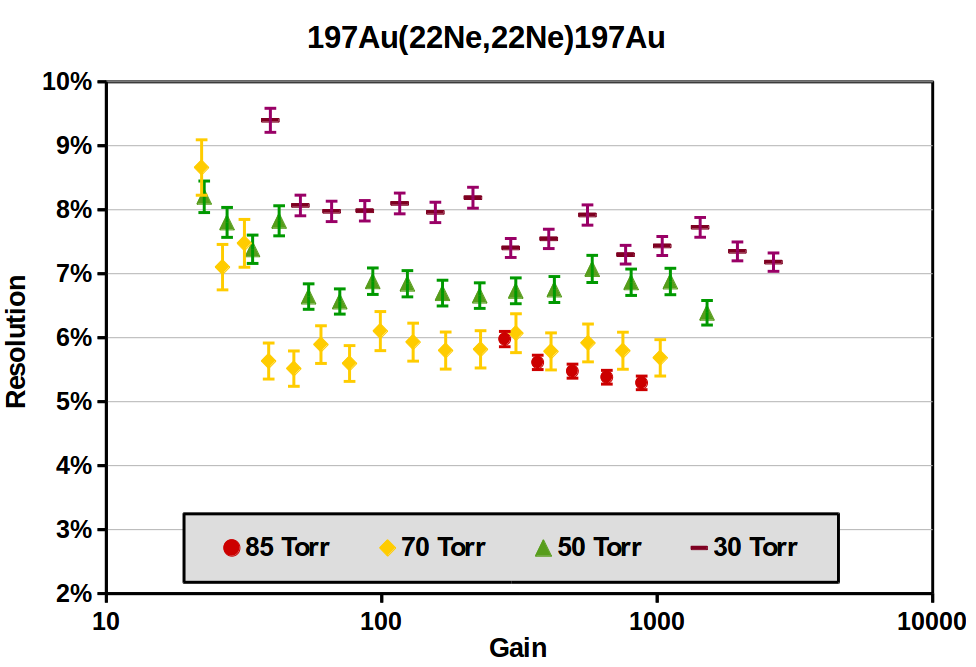}
	\caption{Resolution variation with micromegas gain for pad R1-C4 for pressures of 85, 70, 50 and 30 Torr.}
	\label{fig:4.24}
\end{figure}

The resolution variation across the Micromegas anode pads was also determined and can be
seen in Fig. \ref{fig:ResAll-Al50}, for the case of $^{27}$Al nuclei and 50 Torr pressure. The pads in row 1 generally have better resolution then the ones in the other three rows. This is due to the fact that beam straggling is less in the gas region of that row than in the later ones. Straggling is also affected by gas pressure and Fig. \ref{fig:4.24} shows how the resolution of pad R1-C4 varies for the case of $^{22}$Ne, for 4 different pressures. As expected, the resolution worsens when the pressure decreases and the energy straggling increases.

For Micromegas, the overall range of values for the energy loss pad resolution, taking into account the different nuclei and settings is 5-11\%. This is to be compared to the \textbf{dE1} resolutions of 13-20\% for the original detector. Micromegas is definitely the better option.
\section{Charge Sharing}
When the beam is tightly collimated, it is simple to make sure that only one column of pads detects the particles. Typically, for nuclear physics experiments with the MDM-Oxford, the collimation mask is much wider, specifically 4\degree{} wide (lab system). Additionally, the targets used produce a variety of reaction products. As such, the particle paths cover the entire focal plane.

\begin{figure}[!ht]
	\begin{center}
		\subfloat[]{%
			\includegraphics[width=0.5\linewidth]{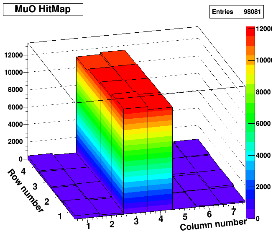}
		}%
		\subfloat[]{%
			\includegraphics[width=0.4\linewidth]{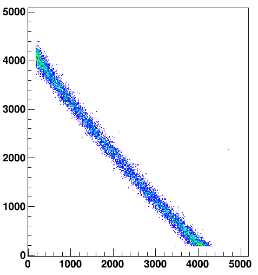}
		}\\
		\subfloat[]{%
			\includegraphics[width=0.45\linewidth]{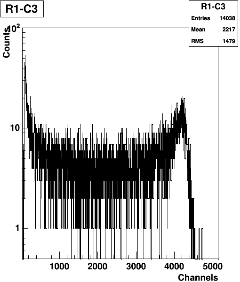}
		}%
		\subfloat[]{%
			\includegraphics[width=0.45\linewidth]{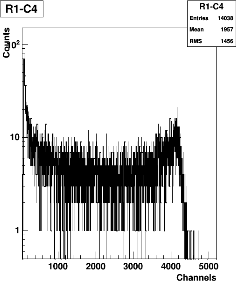}
		}        
	\end{center}
	\caption{(a) 3-D hitmap showing the path of the beam. (b) 2-D histogram showing data from R1-C3 on Y-axis and data from R1-C4 on X-axis. (c) Histogram showing raw data for pad R1-C3. (d) Histogram showing raw data for pad R1-C4.}
	\label{fig:ChShare}
\end{figure}

For the Micromegas anode, specifically, this means that often ionization occurs in such a way that the resulting avalanche curtain cloud can split between adjacent pads. Fig. \ref{fig:ChShare} shows an example of charge sharing, where a beam of $^{22}$Ne particles was guided through the gas region between columns 3 and 4. Histogram (c) is the 3-D hitmap of the Micromegas anode showing which pads detect a signal. Histogram (b) shows the charge sharing pads in the first row, with R1-C3 on the Y-axis and R1-C4 on the X-axis. The remaining histograms were placed next to their respective axes to show the individual pad responses.
%\FloatBarrier
%

In order to obtain an accurate measure of the energy loss of the ionizing particle, the amplified charge needs to be reconstructed properly from these separate individual signals. However, there are two issues that complicate the reconstruction process. The first is that the gain may not be completely uniform across all the pads. The second problem is the danger of losing part of the signal in some cases. For example, if the charge sharing is largely uneven, it is possible that one part of the signal is so small as to register below the ADC or discriminator thresholds. In that case, the reconstructed signal amplitude is smaller than it should be and could lead to misinterpretation of the obtained data.

The non-uniformity issue was solved by gain-matching the pads. This procedure involves sweeping the beam across the anode. The tightly collimated beam loses approximately the same energy in each column and can be used to relate the pads to each other in each row. Any differences in path length due to the entrance angle into the detector are small enough to be negligible.
\begin{figure}[!ht]
	\begin{center}
		\subfloat[]{%
			\includegraphics[width=0.48\linewidth]{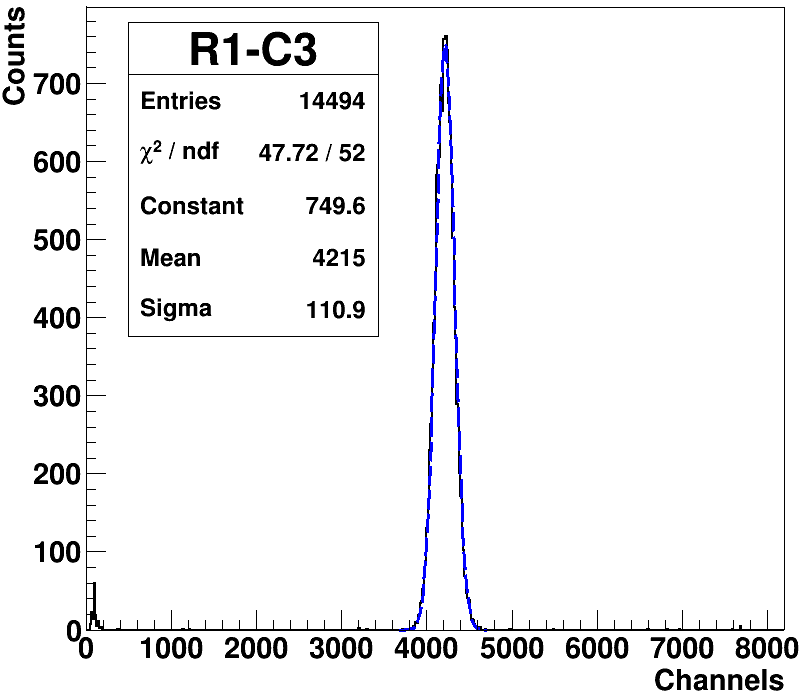}
		}%
		\subfloat[]{%
			\includegraphics[width=0.49\linewidth]{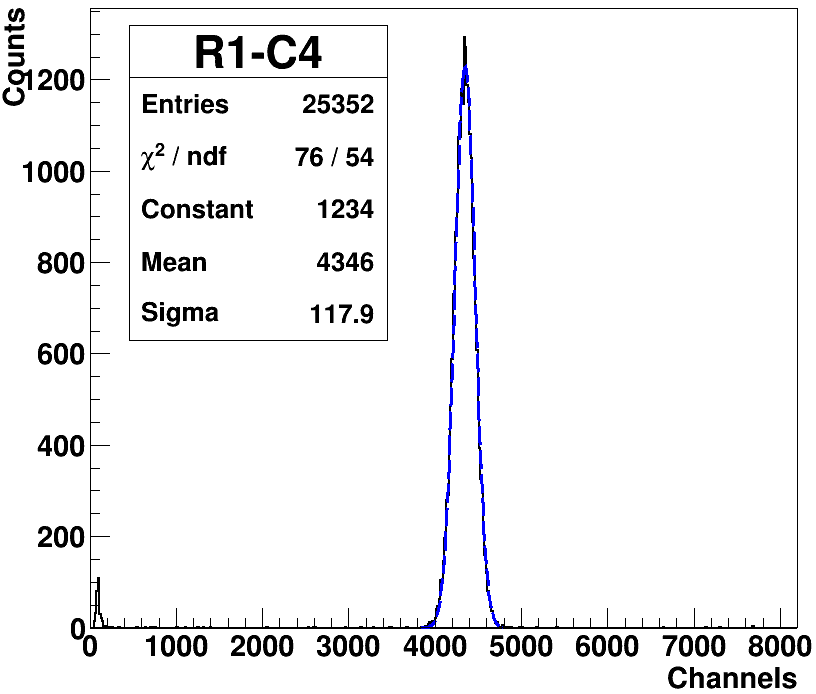}
		}\\
		\subfloat[]{%
			\includegraphics[width=0.48\linewidth]{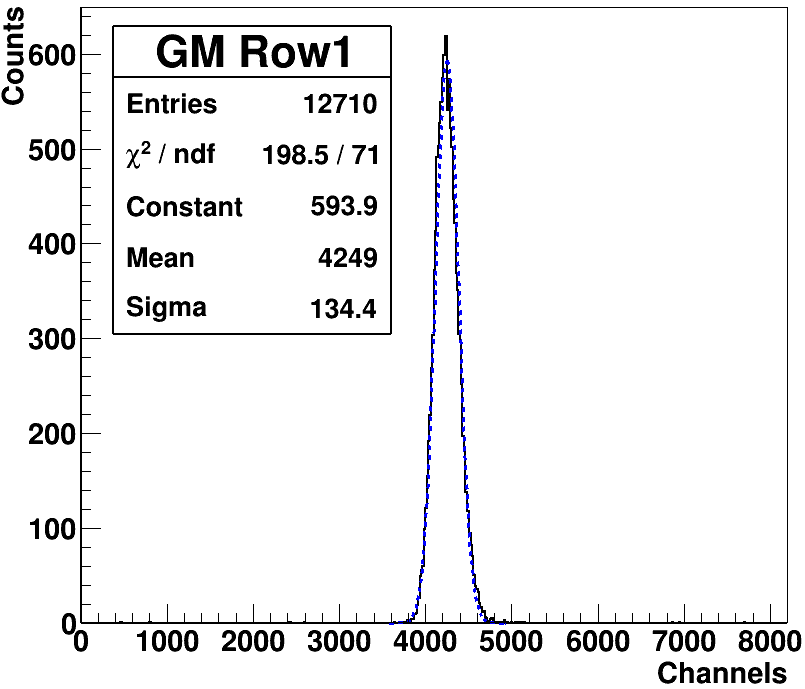}
		}%
		\subfloat[]{%
			\includegraphics[width=0.49\linewidth]{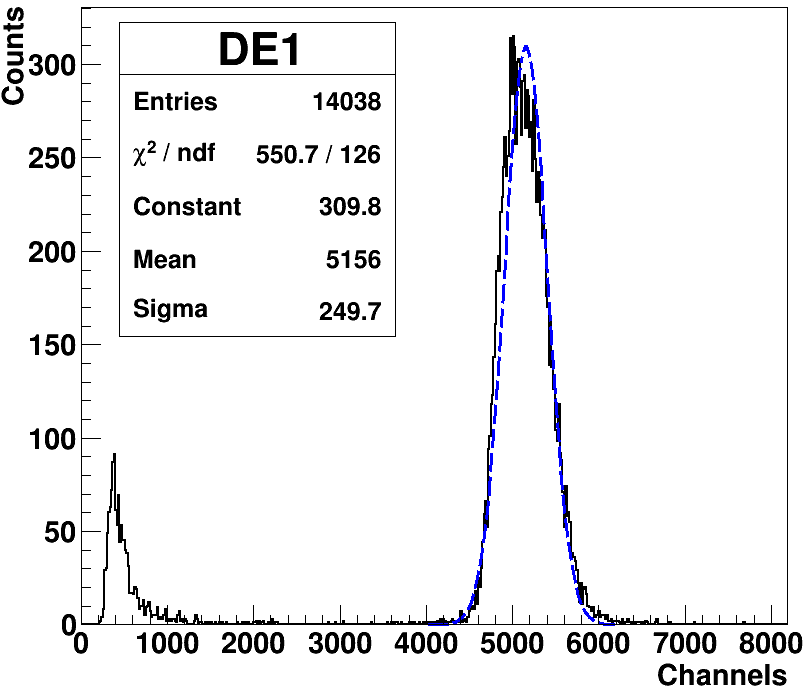}
		}        
	\end{center}
	\caption{(a,b) Energy histograms for pads R1-C3 and R1-C4. (c) Histogram showing the reconstructed energy. (d) Histogram showing dE1, the energy loss signal for the ionization chamber.}
	\label{fig:ReconHist}
\end{figure}

The second issue is more difficult to resolve. The biggest obstacle is the electronic noise. In order to reduce the amount of signal lost, the system noise must be as small as possible. Unfortunately, some of the noise contributions come from the elements in the beam-line, like the power supplies for the magnets and the vibrations caused by the vacuum cryo-pumps. It was not possible to fully isolate the detector from those noise sources.

However, if the noise can be minimized the effects of the lost data are less pronounced. Furthermore, for the purpose of particle identification the significant improvement in resolution compensates for these defects in reconstruction.

Taking all these into account, an example of the quality or efficiency of the reconstruction process can be seen in Fig. \ref{fig:ReconHist}. The top plots show the response of pads R1-C3 (resolution $\approx$ 6.2\%) and R1-C4 (resolution $\approx$ 6.4\%), when there is no charge sharing.

The bottom-left histogram, (c), shows the reconstructed peak when the beam passes between the two pads. As expected, the energy resolution is slightly worse, at 6.9\%, and the peak exhibits a small tail on the high energy side. The bottom-right histogram, (d), shows the ionization chamber response, dE1, which is similar in shape but the resolution is significantly worse, at 11.4\%. The number of counts under these two peaks differs by less than 0.1\%.

As such, the reconstruction method was considered successful and was used in the following analysis and in later experiments.
\section{Calculating the Total Anode Energy Loss}
The first step in obtaining the total Micromegas anode energy loss is to gain-match the pads as explained above. The second step is to determine the multiplicity of an event for each row. Since a particle can either ‘hit’ one pad in a row or ‘hit’ between two neighboring pads in the same row, the multiplicity per row should only be 1 or 2 with adjacent pads. Events not satisfying these criteria for each row are considered non- physical. Under these circumstances, the energy detected by each row is determined from the sum of the individual, gain-matched, responses of the pads in each row. The final step in obtaining the total energy is to calculate the sum of the 4 rows.
 
During testing, several different methods were tried for `summing' the 4 row energies. The sum of the gain-matched row energies produced a total resolution better by almost a factor of 2 than the single pad resolution. This can be easily understood from the fact that the initial number of electrons in all four cases (i.e. for each of the four rows) is roughly the same, therefore their sum is four times larger and correspondingly the relative resolution is $\sqrt{4}=2$ times better, as it is dominated by statistics in the first stage (ionization). Averaging (arithmetic and geometric) was attempted as well and produced similar results to this sum.

A comparative analysis was done for all the scattered beams used as a function of pressure. The results are given in Tab. \ref{tab:4.Resl}. The label E$_{\text{Micromegas}}$ represents the sum of the rows described above. The energy loss resolutions for \textbf{dE1} and the first Micromegas row are also given for comparison purposes.
\begin{table}[ht!]
	\centering
	\begin{tabular}{| c | c | c | c |}
		\hline \hline
		Beam &  R$_\textbf{dE1}$ & R$_\textbf{Row}$ & R$_{\textbf{E}_{\text{Micromegas}}}$ \\\relax
		[Pressure in Torr] & [\%] & [\%] & [\%] \\
		\hline
		$^{16}$O [100] & 8.7 $\pm$ 0.3 & 5.3 $\pm$ 0.2 & 2.9 $\pm$ 0.1 \\ \hline
		$^{22}$Ne [30] & 12.2 $\pm$ 0.3 & 7.4 $\pm$  0.2 & 4.7 $\pm$ 0.1 \\ \hline
		$^{22}$Ne [50] & 10.9 $\pm$ 0.2 & 6.6 $\pm$ 0.1 & 3.7 $\pm$ 0.1 \\ \hline
		$^{22}$Ne [70] & 9.8 $\pm$ 0.2 & 5.6 $\pm$ 0.1 & 3.2 $\pm$ 0.1 \\ \hline
		$^{22}$Ne [100] & 10.9 $\pm$ 0.2 & 6.5 $\pm$ 0.1 & 4.3 $\pm$ 0.1 \\ \hline
		$^{26}$Mg [30] & 7.5 $\pm$ 0.1 & 7.7 $\pm$ 0.1 & 4.4 $\pm$ 0.1 \\ \hline
		$^{27}$Al [50] & 5.3 $\pm$ 0.1 & 6.5 $\pm$ 0.1 & 4.0 $\pm$ 0.1 \\ \hline
		$^{28}$Si [30] & 7.9 $\pm$ 0.1 & 8.8 $\pm$ 0.1 & 6.1 $\pm$ 0.1 \\ \hline
		$^{28}$Si [70] & 6.1 $\pm$ 0.1 & 7.5 $\pm$ 0.1 & 4.9 $\pm$ 0.1 \\ \hline
		$^{32}$S  [30] & 14.9 $\pm$ 0.3 & 11.0 $\pm$ 0.2 & 6.9 $\pm$ 0.1 \\ \hline
		$^{32}$S  [50] & 7.9 $\pm$ 0.2 & 9.2 $\pm$ 0.2 & 5.6 $\pm$ 0.1 \\
		\hline \hline
	\end{tabular}
	\caption{Energy loss resolutions for different detection elements and for the different test beams used.}
	\label{tab:4.Resl}
\end{table}
\begin{figure}[!ht]
	\centering
	\includegraphics[width=\linewidth]{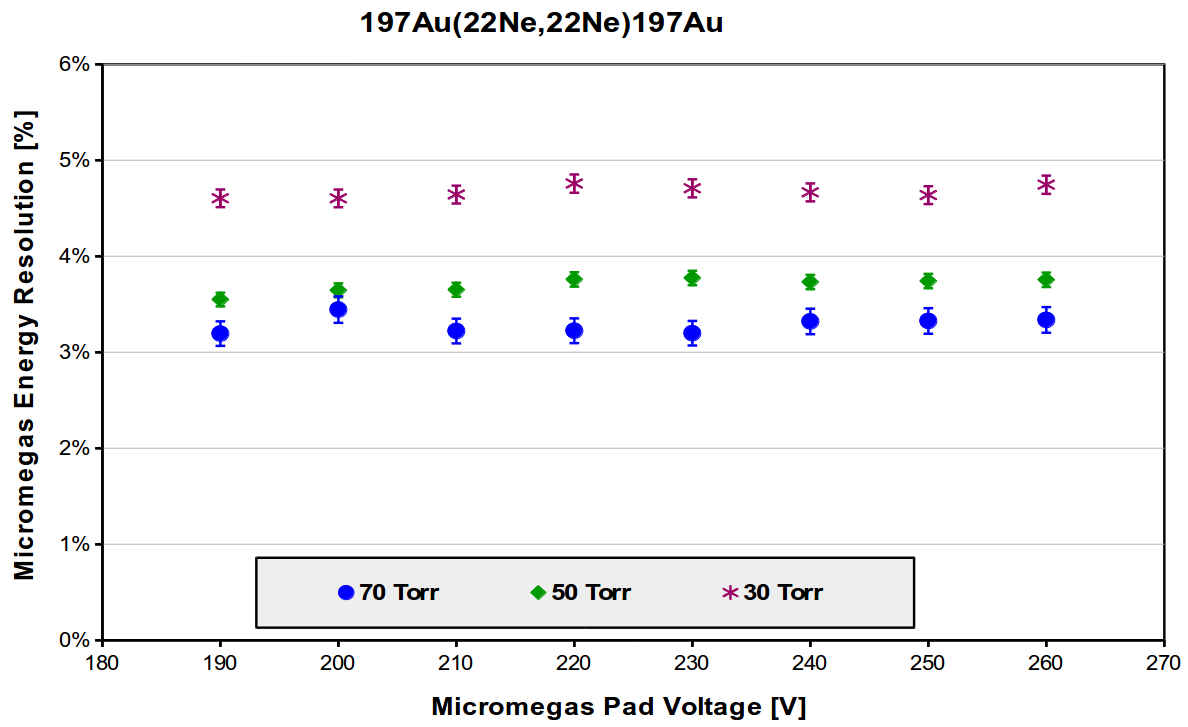}
	\caption{Total energy resolution (E$_{\text{Micromegas}}$) for different Micromegas pad voltages and 3 different pressures (colors labeled in the legend) for $^{22}$Ne beam on Au target.}
	\label{fig:TotalRes}
\end{figure}

\begin{figure*}[!ht]
	\begin{center}
		\subfloat[]{%
			\includegraphics[width=0.45\textwidth]{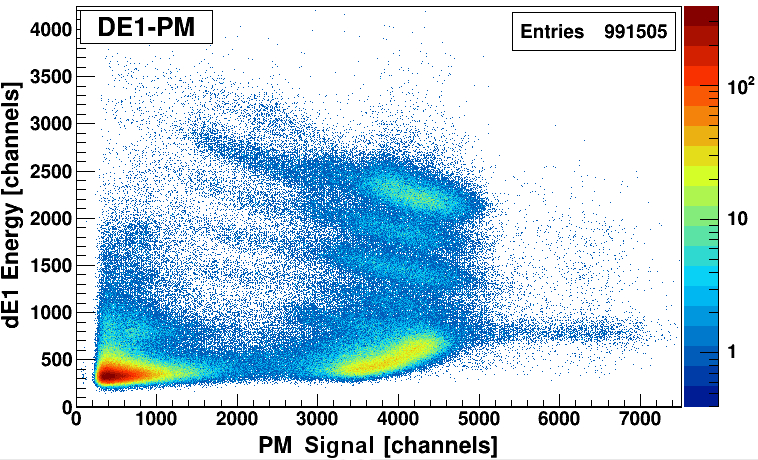}
		}%
		\hspace{0.25 cm}
		\subfloat[]{%
			\includegraphics[width=0.45\textwidth]{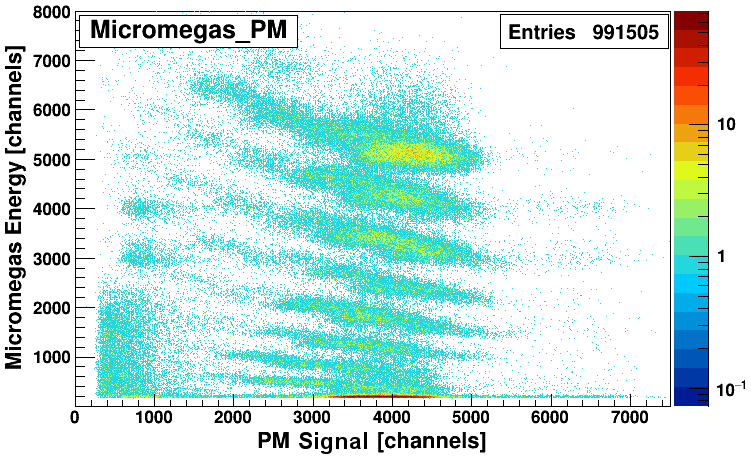}
		}
	\end{center}
	\caption{PID plots for the reaction $^{22}$Ne(12 MeV/A) + $^{13}$C at pressure=30 Torr. (a) PID histogram representing the energy loss \textbf{dE1} versus residual energy. (b) PID histogram representing Micromegas energy loss versus same residual energy.}
	\label{fig:dE1-MuO}
\end{figure*}

It can be seen from Tab. \ref{tab:4.Resl} that given the relatively high gain coupled with a good signal to noise ratio, the upgrade allows a multi-sample of the energy loss that yields a significant improvement in the energy resolution. Fig. \ref{fig:TotalRes} shows the total Micromegas energy (E$_{\text{Micromegas}}$) resolution for different bias voltages and pressures. Comparing these with the numbers in Fig. \ref{fig:4.24} shows that the improvement in resolution by a factor of $\sim$2 due to multi-sampling holds for a wide range of bias voltages. This in turn means that the choice of pad bias doesn't affect the total anode energy as much as it does individual pads, therefore allowing a larger optimal operational range for the Micromegas. Finally, Fig. \ref{fig:dE1-MuO} shows distinctly the
difference between the ionization chamber and the Micromegas upgrade.

The two particle identification plots were recorded at the same time in identical conditions: cocktail products from the reaction $^{22}$Ne+$^{13}$C at a kinematic angular range of 7-11\degree{} lab and 30 Torr pressure in the Oxford detector.
\section{Conclusions}

We have introduced and studied an upgrade of the existing ``Oxford detector'' used in the focal plane of the MDM spectrometer at the Cyclotron Institute, Texas A\&M University. The purpose was to enhance the resolution of the particle identification. The upgrade consists of a system of 4 rows x 7 columns = 28 pads with Micromegas technology to amplify the energy loss signal from the ionization chamber part of the gas detector. It was placed in the second half of the existing detector, while keeping the first part of it with the existing solution. We show herein that the Micromegas operates well even at the lower gas pressures (30 -- 200 Torr), an important regime since the Oxford detector is used for heavy ions at moderate energies (10--20 MeV/nucleon), which require operation at these low pressures.

Up to now Micromegas were used at pressures around 1 atm. With moderate bias voltages of $\sim$280 V, the Micromegas could be run to obtain energy loss resolutions 2 to 3 times better than the previous method, thus extending the particle identification capabilities well into the A=40 region. We proved that the system remains linear for a wide range of energy losses, that inter-pad gaps lead to minor losses, however, the position reconstruction of the detector is not affected. While the increased number of pads complicates the acquisition and the analysis of the data (28 signals instead of one), the advantage is worthy and easily handled with today's technologies.

The modified detector was tested with count rates on the order of tens of kHz and found to be performing within the above stated parameters. A limit of 50 kHz was determined and attributed to the fragility of the wires used for the avalanche counters. The Micromegas component showed no problems with the increased rate. A separate study was performed on the performance of Micromegas as a function of rates and the extracted time resolution when compared with the PM.

Further improvement could come from padding the whole anode of the detector with Micromegas and using raytrace reconstruction in particle identification to allow comprehensive corrections which should improve even further the resolution. In our tests so far, such corrections were unnecessary as the differences in path length inside the Oxford detector were less than 1\%. However, calculations show these differences increasing for heavier nuclei, higher reaction angles and increased acceptance of the detector.

\section{Acknowledgements}

The work done for this study was funded in part by the US DOE under Grants DE-FG02-93ER40773 and DE-NA0001785 and by Project PNIII/P5/P5.2 nr. 02/FAIR-RO of the Romanian Ministry of Research.

\section*{References}

\bibliography{mybibfile}
\end{document}